\documentclass{PoS}

\usepackage{amsmath,amssymb,url}
\usepackage{graphicx,xcolor}
\usepackage{braket}
\usepackage{fancyvrb}

\usepackage{amsfonts}
\usepackage{slashed}
\usepackage{mathrsfs }
\usepackage{bbold}
\usepackage{mathtools}
\usepackage{caption}
\usepackage{subcaption}
\usepackage[makeroom]{cancel}
\usepackage{amsthm}
\usepackage{physics}
\usepackage{bm}
\usepackage{simplewick}

\renewcommand{\comm}[2]{\left[ #1, #2 \right]}

\title{Lattice Computation of the Ghost Propagator in Linear Covariant Gauges}

\ShortTitle{Lattice LCG Ghost Propagator}

\author{Attilio Cucchieri,$^a$ David Dudal,$^{bc}$ Tereza Mendes,$^a$
Orlando Oliveira,$^d$ Martin Roelfs,$^{b}$ and \speaker{Paulo J. Silva}$^{,d}$\\
\llap{$^a$} Instituto de F\'{i}sica de S\~{a}o Carlos, Universidade de S\~{a}o Paulo, C.P. 369, 13560-970 S\~{a}o Carlos, SP, Brazil\\
\llap{$^b$} KU Leuven Kulak, Department of Physics, Etienne Sabbelaan 53 bus 7657, 8500 Kortrijk, Belgium \\
\llap{$^c$} Ghent University, Department of Physics and Astronomy, Krijgslaan 281-S9, 9000 Gent, Belgium\\
\llap{$^d$} CFisUC, Departamento de F\'{i}sica, Universidade de Coimbra, 3004-516 Coimbra, Portugal\\
E-mail: \email{attilio@ifsc.usp.br}, \email{david.dudal@kuleuven.be},
\email{mendes@ifsc.usp.br}, \email{orlando@fis.uc.pt}, \email{martin.roelfs@kuleuven.be}, \email{psilva@uc.pt}}

\abstract{We discuss the subtleties concerning the lattice computation of the ghost propagator in linear covariant gauges, and present preliminary numerical results.}

\FullConference{The 36th Annual International Symposium on Lattice Field Theory - LATTICE2018\\
		22-28 July, 2018\\
		Michigan State University, East Lansing, Michigan, USA.}

\begin{document}

\section{Introduction and Motivation}

The infrared behaviour of Green's functions of Yang-Mills theory has been the subject of many studies in recent years. The relevance of such studies is rooted in the information on the non-perturbative phenomena encoded in the propagators of fundamental fields  of QCD. In particular, gluon and ghost propagators encode information about confinement. While most of the studies of QCD propagators are done in Landau gauge $\partial_{\mu} A_{\mu}(x)=0$ \cite{Alkofer2001, Greensite2011, Vandersickel2012, Maas2013, Binosi2009, Boucaud2012}, we would like to go beyond this gauge to understand the gauge dependent properties of QCD propagators.

Here we consider the linear covariant gauge (LCG), defined by  $\partial_{\mu} A_{\mu}(x)=\Lambda(x)$, where $\Lambda(x)=\Lambda^{a}(x) t^{a}$ are matrices belonging to the SU(N) Lie algebra, and $\Lambda^{a}(x)$ are random real numbers, Gaussian distributed around zero with a variance $\xi$.

The LCG gluon propagator has already been studied on the lattice by some authors \cite{Giusti2001, Cucchieri2009, Bicudo2015, CucchieriPRL2009}.
In this paper we report a lattice calculation of the LCG ghost propagator --- see also \cite{Cucchieri2018} for a recent report.

\section{Landau and LCG ghost propagator on the lattice}

On the lattice, the Landau gauge is defined through the numerical optimization, along the gauge orbit, of the gauge fixing functional
  \begin{equation} 
   F^{Landau}(U^{g}) =  - \sum_{x, \mu} \Re  \tr \left[
   \; g(x) \, U_{\mu}(x) \, g^{\dagger}(x+  \hat{\mu}) \right]. \label{flandau}
  \end{equation}
From the first variation of eq. (\ref{flandau}) one gets a lattice version of the  Landau gauge condition $\partial_{\mu} A^{a}_{\mu}=0$, whereas the second variation defines the symmetric matrix
	\begin{eqnarray}
		M^{ab}_{xy} &=& \sum_\mu \Re \tr \bqty{\acomm{t^a}{t^b}\pqty{U_\mu(x) + U_\mu(x - \hat{\mu})}} \delta_{xy} \nonumber \\
		&-& 2 \sum_\mu \Re \tr \bqty{t^b t^a U_\mu(x)} \delta_{x+\hat{\mu},y} - 2 \sum_\mu \Re \tr \bqty{t^a t^b U_\mu(x - \hat{\mu})} \delta_{x-\hat{\mu},y}. \label{FPLandau}
	\end{eqnarray}
At some minimum of the functional (\ref{flandau}), $M^{ab}_{xy}$ is positive semi-definite. One can show that (\ref{FPLandau}) is a suitable discretization of the continuum operator  $-\frac{1}{2}\left(\partial_{\mu} D^{ab}_{\mu}+D^{ab}_{\mu} \partial_{\mu}\right)$, which in the Landau gauge is equal to  $ -\partial_{\mu} D^{ab}_{\mu}$, i.e.  the usual Faddeev-Popov (FP) operator. The lattice approach to compute the Landau gauge ghost propagator consists in inverting the matrix described by eq. (\ref{FPLandau}). Since $M^{ab}_{xy}$ is symmetric and positive semi-definite, the Conjugate Gradient method can be used to perform such inversion.

Similarly, the linear covariant gauge can be defined on the lattice through the numerical optimization of the gauge fixing functional \cite{CucchieriPRL2009}
  \begin{equation} 
   F^{LCG}(U^{g};g) =  F^{Landau}(U^{g}) + \Re \tr \sum_x \left[ i g(x) \Lambda(x) \right]. \label{flcg}
  \end{equation}
The first variation defines the lattice analogue of the  LCG condition in the continuum, whereas the second variation defines the same symmetric matrix, eq. (\ref{FPLandau}), as in Landau gauge. However, in the LCG case,  eq. (\ref{FPLandau}) is not a suitable discretization of the continuum FP operator. 

A suitable lattice discretization of the LCG FP operator, with the correct continuum limit, can be found by defining the lattice operators 
\begin{eqnarray}
		\left[M^{+}\right]^{ab}_{xy} &=& M_{xy}^{ab} + \left[\Delta M\right]^{ab}_{xy}  \label{Mplus} \\
		\left[M^{-}\right]^{ab}_{xy} &=&  M_{xy}^{ab} - \left[\Delta M\right]^{ab}_{xy}   \label{Mminus}
\end{eqnarray}
where 
\begin{equation}
\left[\Delta M\right]^{ab}_{xy}  = \Re \tr \sum_\mu \bqty{\comm{t^a}{t^b} \pqty{U_\mu(x) - U_\mu(x - \hat{\mu}) } } \delta_{xy} \label{delta_m}.
\end{equation}
The matrices $M^{+}$ and $M^{-}$ are suitable discretizations of the continuum operators $-\partial_\mu D_\mu$ and $-D_\mu \partial_\mu$ respectively.  Note that $M$, $M^{+}$ and $M^{-}$ can not be distinguished as quadratic forms, in the sense that 
\begin{equation}
\omega^a(x) \left[\Delta M\right]^{ab}_{xy} \omega^b(y) = \omega^a(x) f_{abc} \Re \tr [i t^c \pqty{U_\mu(x) - U_\mu(x - \hat{\mu}) } ] \omega^b(y) = 0 , \label{nullm}
\end{equation}
due to the antisymmetry of the structure constants $f_{abc}$.

\section{Results}

The matrix $M^{+}$ provides a suitable lattice discretization of the continuum FP operator, enabling a lattice computation of the LCG ghost propagator. Since $M^{+}$ is a real non-symmetric matrix, it can not be inverted using Conjugate Gradient method (as in Landau gauge) and, therefore, we rely on  the Generalized Conjugate Residual method, described e.g. in \cite{Saad}. To avoid possible zero modes\footnote{Note that, in the LCG case, constant vectors are not zero modes of $M^{+}$.} of $M^+$ , we solve the system \cite{Suman1996} 
\begin{displaymath}
M^+ M^+ X = M^+ b 
\end{displaymath}
that, for performance purposes,   is solved in two steps
\begin{eqnarray}
M^+ Y &=& M^+ b;  \nonumber \\
M^+ X &=& Y.  \nonumber
\end{eqnarray}

\begin{figure}
\vspace*{1cm}
\begin{center}
\includegraphics[width=0.7\textwidth]{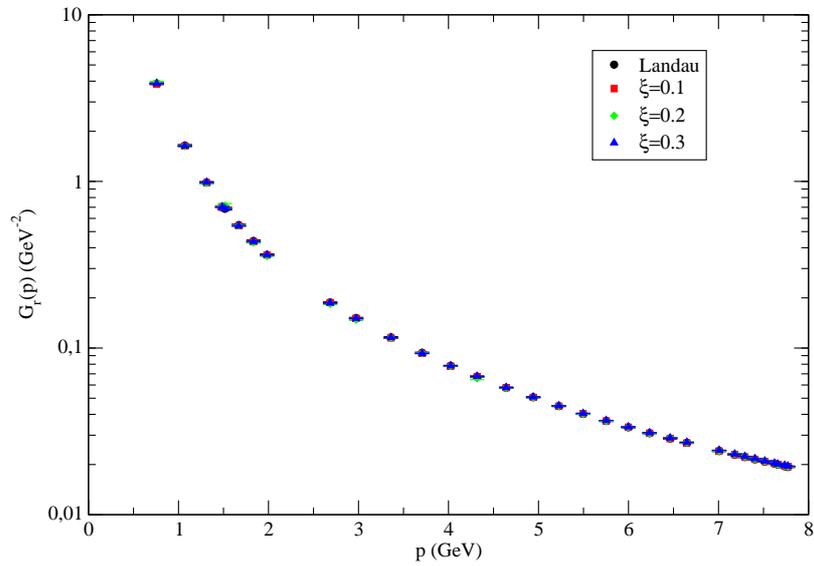}
\end{center}
\caption{Landau and LCG ghost propagators ($\xi\in\{0.1, 0.2, 0.3\}$) for a $16^4$ lattice volume.}
\label{fig16}
\end{figure}

\begin{figure}
\vspace*{1cm}
\begin{center}
\includegraphics[width=0.7\textwidth]{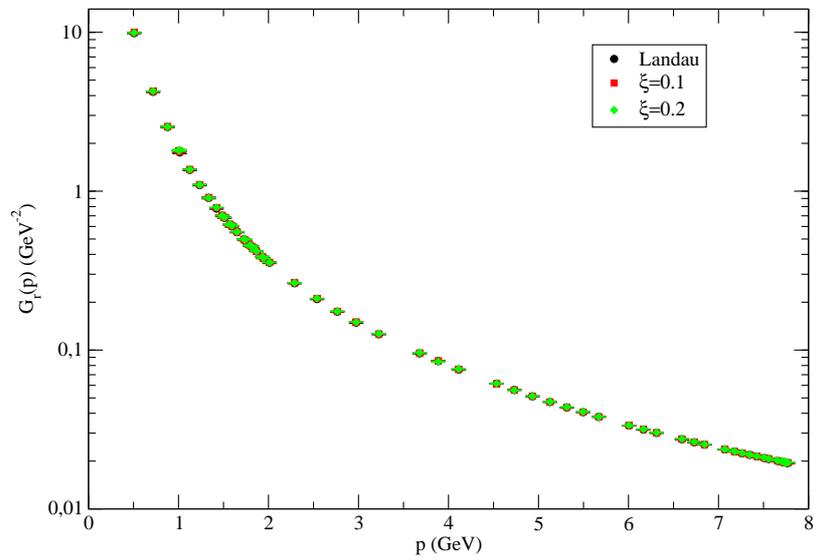}
\end{center}
\caption{Landau and LCG ghost propagators ($\xi\in\{0.1, 0.2\}$) for a $24^4$ lattice volume.}
\label{fig24}
\end{figure}

In Figures \ref{fig16} and \ref{fig24} we report our results for the LCG ghost propagator, evaluated using a point source for the inversion. We considered SU(3) pure gauge simulations using the Wilson action at $\beta=6.0$, which corresponds to a lattice spacing $a\sim 0.102$ fm.  For $16^4$ and $24^4$ lattice volumes, we have generated 100 thermalized gauge configurations, and 20 sets of Gaussian-distributed $\{\Lambda(x)\}$ matrices for each configuration. We compare with the Landau gauge ghost propagator, computed from the same set of configurations. No clear difference between Landau and LCG data is observed in the plots.

\section{Conclusion}

We discussed an approach to compute the LCG ghost propagator on the lattice, and presented numerical results for small lattice volumes. LCG lattice data is in agreement with Landau gauge results. Similar results have been obtained using SU(2) pure gauge simulations \cite{Cucchieri2018, conf13}.

\section*{Acknowledgements}

A.C. and T. M. acknowledge partial support from CNPq.
A.C. also acknowledges partial support from FAPESP (grant
\# 16/22732-1). The research of D.D. and M.R. is supported
by KU Leuven IF project C14/16/067. The authors O.O. and
P.J.S. acknowledge the Laboratory for Advanced Computing
at University of Coimbra (http://www.uc.pt/lca)
for providing access to the HPC computing resource
Navigator. O. Oliveira and
P. J. Silva acknowledge financial support from FCT Portugal under contract with
reference UID/FIS/04564/2016. P.J.S. acknowledges support by Funda\c{c}\~ao 
Luso-Americana para o Desenvolvimento (FLAD Proj. 155/2018), 
and  by Funda\c{c}\~ao para a Ci\^encia e a Tecnologia (FCT) under contracts
SFRH/BPD/40998/2007 and SFRH/BPD/109971/2015.
The SU(3) lattice simulations were done using Chroma \cite{Edwards2005} and
PFFT \cite{Pippig2013} libraries.

\end{document}